\documentclass[11pt]{article}
\usepackage{graphicx,color}
\usepackage{makeidx}
\usepackage{float}
\usepackage{hyperref}
\usepackage{setspace}
\usepackage{algorithm} 
\usepackage{algorithmic} 
\usepackage{booktabs} 

\usepackage{amsmath}
\usepackage{subfig}
\usepackage{framed}
\def\qed{\rule[0mm]{.6em}{.6em}}

\begin{document}

\bibliographystyle{plain}
\setcounter{secnumdepth}{3}
\setcounter{tocdepth}{3}
\newtheorem{Lemma}{Lemma}[section]
\newtheorem{Theorem}{Theorem}[section]
\newtheorem{Corollary}{Corollary}[section]
\newenvironment{Proof}{{\bf Proof }}{\qed}
\newenvironment{Algorithm}{\medskip\small\mbox{}\\\begin{tabbing}\tt}{\end{tabbing}\normalsize}
\newenvironment{Subheading}{\medskip\mbox{}\\ \noindent \bf}{\rm \smallskip\\}
\newcommand{\len}{{\tt{Len}}}
\newcommand{\str}{{\tt{Str}}}
\newcommand{\II}{{\cal I}}
\newcommand{\MM}{{\cal M}}
\newcommand{\DD}{{\cal D}}
\newcommand{\CC}{{\cal C}}
\newcommand{\BB}{{\cal B}}
\newcommand{\KK}{{\cal K}}
\newcommand{\WW}{{\cal W}}
\newcommand{\XX}{{\cal X}}
\newcommand{\YY}{{\cal Y}}
\newcommand{\LL}{{\cal L}}
\newcommand{\ZZ}{{\cal Z}}
\newcommand{\UU}{{\cal U}}
\newcommand{\VV}{{\cal V}}
\newcommand{\TT}{{\cal T}}
\newcommand{\SSS}{{\cal S}}
\newcommand{\RR}{{\cal R}}
\newcommand{\HH}{{\cal H}}
\newcommand{\CA}{{\cal A}}
\newcommand{\PP}{{\cal P}}
\newcommand{\FF}{{\cal F}}
\newcommand{\ZM}{{\cal M}}
\newcommand{\F}{{\cal F}}
\newcommand{\Q}{{\cal P}}
\newcommand{\N}{{\cal N}}
\newcommand{\lra}{\leftrightarrow}
\newcommand{\la}{\leftarrow}
\newcommand{\ua}{\uparrow}
\newcommand{\nwa}{\nwarrow}
\newcommand{\da}{\nwarrow}

\newcommand{\nn}{\noindent}
\newcommand{\mm}{\medskip}
\newcommand{\Caption}[1]{\caption{\small #1}}


\begin{center}
\Large{Matrix Chain Multiplication and Polygon Triangulation Revisited and Generalized}\footnote{Written in 2017. 
Citation \cite{SCHWEI2019} and comment added in 2021. This work was supported by NSF grant 1528234.}

\medskip

Thong Le and Dan Gusfield\\
\medskip
U.C. Davis, Computer Science

\today
\end{center}

\begin{abstract}
The {\it matrix-chain multiplication} problem is a classic problem that is widely taught 
to illustrate dynamic programming. The textbook solution runs in $\theta(n^3)$ time.
However, there is a complex $O(n \log n)$-time method \cite{HU82}, based on triangulating convex polygons, 
and a description without proofs or implementation detail, of a much simpler $O(n^2)$-time method \cite{YAO82}. There is also a 
linear-time approximation algorithm with a small worst-case error bound \cite{HU-SHING1981}. In this paper,
we make five contributions both to theory and pedagogy:
1) We simplify the approach in \cite{YAO82}, and provide complete, correct proofs and implementation details, to
establish the $O(n^2)$-time bound.  We believe that this exposition is simple enough for classroom use.
2) We extend the $O(n^2)$-time bound 
to a natural class of polygon-triangulation problems that generalizes the original polygon-triangulation problem in \cite{HU82}.
3) We show that the worst-case running time of the method in \cite{YAO82}, and
of our version, is $\Theta(n^2)$.
4) We show that in a natural variant of the original polygon-triangulation problem,
the approximation method of \cite{HU-SHING1981} does not achieve the same error bound, but does achieve an
error bound about twice the original bound.
5) We detail empirical testing, showing that on random data our variant runs in $\Theta(n \log n)$ time,
while the approach in \cite{YAO82} empirically takes $\Theta(n^2)$ time. Software for these tests is posted on the web.
\end{abstract}

\section{Introduction}
The {\it matrix-chain multiplication} problem is a classic problem that is very widely taught in computer science
algorithms and programming classes, and included in many textbooks (for example \cite{CLR}), and on-line lectures
illustrating the method of dynamic programming.\footnote{A Google advanced search for the exact phrase ``matrix chain multiplication" returns 23,000 hits.} The 
input consists of a sequence of $n$ {\it pairs} of numbers representing the dimensions of $n$ matrices, where for each of the
first $n-1$ matrices, the second dimension of the matrix is equal to the first dimension of the next matrix. 
Then, the product of the matrices must be computed; but since matrix 
multiplication is {\it associative}, different sequences of matrix multiplications are possible, with a different
number of individual, {\it scalar} multiplications.
The problem is to choose a sequence of matrix multiplications, 
specified by a nested parenthesization of the input matrices, 
to {\it minimize} the number of scalar multiplications.  The classic dynamic programming
solution to find an optimal parenthesizing takes $\Theta(n^3)$ time \cite{CLR}. 

It is well known \cite{HU82,HU-SHING2002} that the matrix-chain multiplication problem can be reduced in $O(n)$ time
to a problem of optimally triangulating a node-weighted {\it polygon} with $n$ nodes, adding exactly $n-3$ internal 
edges,
partitioning the polygon into exactly $n-2$ triangles.  The weight of a triangulation is
the {\it sum} of the weights of the $n-2$ triangles; and the weight of a triangle is the {\it product} of the
weights of its three nodes. This {\it triangle-weight function} is called
{\it multiplicative}.  Based on this reduction, T.C. Hu and M. T. Shing showed how to solve the matrix-chain multiplication
problem in $O(n \log n)$ time \cite{HU82,Hu1984}.
In \cite{HU-SHING1981} they also establish an $O(n)$-time approximation method for the problem, 
with a tight guaranteed error bound of less than 15.5\%.

The reduction in time from $\Theta(n^3)$ to $O(n \log n)$ is
dramatic and rare, as is a linear-time approximation with such a small error bound.
Unfortunately, the two paper journal exposition \cite{HU82,Hu1984} of the $O(n \log n)$-time result is 35 pages long and omits many details.
The technical report on which those papers are based
is one hundred pages long. Subsequent expositions and variations \cite{RAM94,RAM96}
of the $O(n \log n)$ result have also been incomplete and hard to understand. 
Further, a recent paper \cite{SCHWEI2019} details significant errors in the proofs (in \cite{HU82}) 
of the $O(n \log n)$ result, and several ways that the errors influenced subsequent publications. The centrality
of the matrix-chain problem in computer science education, and the subtlety of correctness proofs, justified
publication of corrected proofs \cite{SCHWEI2019} in a premier journal.

In contrast to the complex $O(n \log n)$ result, F. Yao \cite{YAO82} developed and sketched a solution of the
polygon triangulation problem that runs in $O(n^2)$ time and is much simpler than the $O(n \log n)$-time
result, yet faster than the $O(n^3)$-time algorithm. 
That paper omitted proofs,
citing an incomplete proof in \cite{HU82} for a key point. The paper also omitted 
some implementation details needed to achieve the time bound. 
Later, a variant of the original triangulation problem, where each triangle has weight equal to the {\it sum}, rather than
the {\it product}, 
of weights of its three nodes, 
was shown to be solvable in $O(n \log n)$ time in \cite{RAM94}. This triangle-weight function is called
{\it additive}. 
The exposition in \cite{RAM94} is again quite challenging.

Note that
the weight of a triangulation using the additive triangle-weight function is equivalent to the sum over all the nodes of the polygon,
of the number of triangles the node is part of, times the weight of the node. It is also equivalent to the sum over all the nodes, 
of the {\it degree} of the node (the number of edges it touches in the triangulation) minus one, times the weight of the node.
Thus, triangulating a polygon using the additive triangle-weight function, optimizes
the distribution of node degrees, given the node weights.

\medskip

Our paper makes five contributions both to theory and to algorithmic pedagogy:

1) We simplify the approach of Yao \cite{YAO82}, and provide complete, correct proofs, and implementation details, to
establish an $O(n^2)$-time solution to the original polygon triangulation problem.
We believe that this exposition is simple enough for use in advanced algorithms classes.

2) Our exposition establishes a more general result, that the $O(n^2)$ time bound for the polygon triangulation problem
can be achieved for {\it any} triangle-weight function in a natural class that includes multiplicative and additive weighting.
This achieves for the additive variant (and other variants) of the triangulation problem what
the Yao paper \cite{YAO82} achieves for the multiplicative variant.

3) We show, by construction, that the {\it worst-case} running time of Yao's method, and
of our version, is $\Theta(n^2)$.

4) We show that in the additive-weight variant of the polygon triangulation problem, 
 the linear-time method \cite{HU-SHING1981} with error bound of $15.5\%$ does not work, but does achieve a tight 
error bound of $\frac{1}{3}$.

5) We detail empirical testing, showing that on random data our variant of Yao's method runs in $\Theta(n \log n)$ time,
while Yao's original dynamic-programming approach empirically takes $\Theta(n^2)$ time. These empirical results dramatically illustrate 
the utility of top-down memoization compared to bottom-up dynamic programming.

\subsection{Basic definitions and tools} We let $P(n)$ denote an $n$-sided convex polygon. There is a node at each intersection of consecutive sides
of $P(n)$, hence $P(n)$ has $n$ nodes. Each node $x$ on $P(n)$ has an associated weight, denoted $w(x)$. 
Together, $P(n,w)$ denotes an $n$-sided polygon whose node weights are given by
the function $w$. A {\it triangulation} of $P(n)$ is created by adding straight, {\it internal} edges between pairs of nodes of $P(n)$, so that
no pair of internal edges cross (in the interior of $P(n)$), and no additional edges can be added without violating that condition. 
It is well-known, and easy to show by induction, that any triangulation of $P(n)$ contains exactly $n-3$ internal edges, and creates exactly $n-2$
triangles, each with an empty interior. Hence, a triangulation partitions the interior of $P(n)$ into
$n-2$ triangles.

Let $T(n,w)$ denote a polygon triangulation of a node-weighted $n$-node polygon. 
Let $\{x,y,z\}$ denote three nodes in $P(n)$ that form a triangle 
in $T(n,w)$, with weights $w(x), w(y)$ and $w(z)$ respectively. 
A {\it triangle-weight} function $f(x,y,z)$ assigns a  weight to triangle $\{x,y,z\}$.
In the original (multiplicative) triangle-weight function \cite{HU82,Hu1984},
$f(x,y,z) = w(x) \times w(y) \times w(z)$; and in the additive function \cite{RAM94},
$f(x,y,z) = w(x) + w(y) + w(z)$. 
Here we define a class of triangle-weight functions that contains both of these. We call $f$ {\it monotonic},
if $f(x,y,z) < f(a,b,c)$ whenever $w(x) \leq w(a)$ and $w(y) \leq w(b)$ and $w(z) \leq w(c)$, and at least
one of these relations is {\it strict}. The weight of a triangulation $T(n,w)$ is the {\it sum} of the weights of
the $n-2$ triangles of $T(n,w)$.

Throughout the exposition, we assume that all node weights are distinct. We can achieve this by small perturbations 
of the weights if needed, without changing the minimum weight triangulation. 
Below we show
that the polygonal triangulation problem can be solved in $O(n^2)$ time when the triangle-weight function $f$ is
{\it any} monotonic function. We first generalize Theorem 3 from \cite{HU-SHING2002}. 

\begin{Theorem}
\label{tmonotone}
If the triangle-weight function $f$ is monotonic, then in 
{\it every} optimal polygon triangulation $T(n,w)$ of $P(n,w)$, 
the node of smallest weight must be connected (either by an original side-edge of $P(n)$ or an added internal edge) to the two 
nodes of second and third smallest weights.
\end{Theorem}

The case when $f$ is multiplicative is stated and proved (as Theorem 3) in \cite{HU-SHING2002}.
The proof here of Theorem \ref{tmonotone}
follows that proof, but  uses properties of a monotonic function when 
needed.\footnote{Actually, only the basis, and case 4) are 
different from the proof of Theorem 3 in \cite{HU-SHING2002}, but we include all the cases for completeness.} 

\begin{Proof}
Without loss of generality, we assume the nodes are numbered $v_1, v_2, ..., v_n$ in {\it increasing} order of their weights.
The theorem
will be proved by induction on $n$. It is trivially true for $n = 3$; so consider $n = 4$. The theorem
holds trivially if nodes $v_2$ 
and $v_3$ are neighbors of $v_1$ on the polygon. Assume that $v_2$ is not a neighbor of $v_1$, so the neighbors
of $v_1$ are $v_3$ and $v_4$. There are only two possible triangulations: one containing the edge $(v_3, v_4)$,
and one containing the edge $(v_1,v_2)$. The weights of those two triangulations are $f(v_1,v_3,v_4) + f(v_2,v_3,v_4)$
and $f(v_1,v_2,v_3) + f(v_1,v_2,v_4)$ respectively. By the monotonicity of $f$, and the labeling convention, the
first triangulation has weight strictly greater than the second, in agreement with the theorem. 
A similar argument confirms the base case when $v_3$ is not a neighbor of $v_1$.

Now assume the theorem holds for $n \leq k$, where $k$ is at least four, and consider $n = k+1 > 4$. We will show that the
theorem holds for $n = k+1$.  We first need some definitions
and observations.
A node $u$ is called {\it external} in $T(n,w)$ if $u$ is only adjacent in $T(n,w)$ to its two
neighbors on $P(n,w)$.
It is well-known, and easy to establish by induction on $n$, that any triangulation of an $n$-gon contains
at least two external nodes. 
By definition of a triangulation, no additional internal edges can be added to  $T(n,w)$, so the two neighbors of an external node $u$
must be adjacent in $T(n,w)$ via an internal edge.  It follows that if we remove
node $u$ and its two incident edges from  $T(n,w)$, the result is a triangulation, denoted $T(n,w) - u$, of the $(n-1)$-gon
denoted $P(n,w) - u$.
Further, when $T(n,w)$ is {\it any optimal} triangulation of $P(n,w)$, $T(n,w)-u$ is an optimal
triangulation of $P(n,w) - u$. Conversely, if node $u$ is known to be external, the minimum weight triangulation
of $P(n,w)$ consists of an optimal triangulation of $P(n,w) - u$,
plus the triangle made from $u$ and its two neighbors on $P(n,w)$.

Returning to the inductive proof, let $T(n,w)$ be an optimal triangulation of $P(n,w)$.
We examine four cases to prove the inductive step:

Case 1) One of the external nodes of $T(n,w)$, denoted $u$, is not $v_1$ or $v_2$ or $v_3$, so these nodes are each in $P(n,w) - u$. 
By the induction hypothesis, $v_1$ is adjacent to $v_2$ and $v_3$ in $T(n,w) - u$,
and hence also in $T(n,w)$. 

In the remaining cases, {\it all} of the external nodes are from $\overline{V} = \{v_1, v_2, v_3\}$,
and since there at least two external nodes, it is sufficient to enumerate which {\it pair} of nodes from $\overline{V}$ are external.

Case 2) Nodes $v_2$ and $v_3$ are external. By the same
reasoning as in Case 1, $T(n,w) - v_2$ is an optimal triangulation for $P(n,w) - v_2$, and by the induction hypothesis,
$v_1$ is adjacent to $v_3$ in $T(n,w) - v_2$, and hence also in $T(n,w)$. Similarly, considering $T(n,w) - v_3$, $v_1$ is
adjacent to $v_2$ in $T(n,w)$, so $v_1$ is adjacent to {\it both} $v_2$ and $v_3$ in $T(n,w)$.

Case 3) Nodes $v_1$ and $v_2$ are external. By the same reasoning as above, $T(n,w) - v_2$ is an optimal triangulation
of $P(n,w) - v_2$, and by the inductive hypothesis, $v_1$ is adjacent to $v_3$ and $v_4$ in $T(n,w) - v_2$, and hence also
in $T(n,w)$. But $v_1$ is assumed to be external, and so is only adjacent in $T(n,w)$ to its two neighbors in $P(n,w)$, so 
$v_3$ and $v_4$ must be its two neighbors in $P(n,w)$. By the same reasoning, considering $T(n,w) - v_1$, 
$v_3$ and $v_4$ must also be neighbors of $v_2$ in $P(n,w)$. The only polygon that is consistent with these adjacencies is
$P(4)$, which contradicts the assumption that $k+1 > 4$.

Case 4) Nodes $v_1$ and $v_3$ are external. As above, considering $T(n,w) - v_3$, we conclude that $v_3$ and $v_4$ are 
neighbors of $v_1$ on $P(n,w)$; and considering $T(n,w) - v_1$, we conclude that $v_2$ is adjacent to $v_3$ (and $v_4$).
Then, since $v_3$ is assumed to be external, $v_2$ must be neighbor of $v_3$ on $P(n,w)$.
So the nodes $\{v_4, v_1, v_2, v_3\}$ must be arranged consecutively (in that order) on $P(n,w)$, and since $v_1$ is external in $T(n,w)$,
$T(n,w)$ must contain the internal edge $(v_2, v_4)$.
Therefore, $T(n,w)$ consists of the triangle
$\{v_1,v_2,v_4\}$ plus the optimal triangulation of $P(n,w) - v_1$. 
Now consider a different triangulation, $T'(n,w)$, of $P(n,w)$ consisting of the triangle $\{v_1,v_2,v_3\}$, plus the
optimal triangulation of $P(n,w) - v_2$. By monotonicity, $f(v_1,v_2,v_3) < f(v_1,v_2,v_4)$. Further, 
the optimal triangulation of $P(n,w) - v_2$ has strictly less weight
than the optimal triangulation of $P(n,w) - v_1$. 
To see this, note first that $P(n,w) - v_1$ is a polygon on the
ordered list of nodes $\{v_2, v_4, ... , v_3\}$, while $P(n,w) - v_2$ is a polygon on
the ordered list of nodes $\{v_1, v_4, ..., v_3\}$. So, the optimal triangulations of $P(n,w) - v_1$ induces a triangulation
of $P(n,w) - v_2$, where all the triangles are the same, except for
any triangle, $\{v_2,x,y\}$, containing node
$v_2$. Each of those triangles is changed to $\{v_1,x,y\}$ in the induced triangulation of $P(n,w) - v_2$.
But, by monotonicity, $f(v_1,x,y) < f(v_2,x,y)$, for every such triangle, so the weight of the optimal triangulation of $P(n,w) - v_2$
is strictly less than the weight of the optimal triangulation of $P(n,w) - v_1$. We conclude that $T'(n,w)$ has strictly
less weight than $T(n,w)$, which contradicts the assumption that $T(n,w)$ is an optimal triangulation for $P(n,w)$.
Therefore, Case 4) is not possible, and the inductive step is proved.\footnote{Theorem 3 in 
\cite{HU-SHING2002}
extends the original, weaker, version in \cite{HU82} (stated there as Theorem 1), whose given proof is inadequate to 
establish Theorem 1. This is a subtle mathematical and pedagogical point. Theorem 1 in \cite{HU82} only states
that ``there exists" an optimal triangulation containing $(v_1,v_2)$ and $(v_1,v_3)$. But with that weaker hypothesis,
the inductive step given in \cite{HU82} is inadequate. For example, consider Case 2,
that $v_2$ and $v_3$ are external nodes. The weaker inductive hypothesis establishes that
there is an optimal triangulation of $P(n,w)$ containing $(v_1,v_2)$, and that there is also an optimal triangulation of
$P(n,w)$ containing $(v_1,v_3)$, but it does not establish that there is an optimal triangulation of
$P(n,w)$ containing {\it both} of those edges. Additional argument is then needed. That problem is avoided with the claim
that ``every optimal triangulation has the property" rather than ``there exists an optimal triangulation with the property".
This is a nice example where strengthening the inductive claim makes the proof easier. Yao \cite{YAO82} stated the ``every optimal triangulation"
version, but cited \cite{HU82} for its proof.}
\end{Proof}

The next tool in the development of the $O(n^2)$-time method was
stated In \cite{YAO82} without proof.\footnote{It is stated in \cite{YAO82},
that Corollary \ref{cYAO} follows from Theorem 1 in \cite{HU82}, but we do not understand that, although, as shown here, 
it does follow from the stronger Theorem 3 in \cite{HU-SHING2002}.}

\begin{Corollary}
\label{cYAO}
If the neighbors of $v_1$ in $P(n,w)$ are $v_2$ and $v_3$, then every optimal triangulation $T(n,w)$  either contains
the internal edge $(v_2,v_3)$, or the internal edge $(v_1,v_4)$.
\end{Corollary}

\begin{Proof}
Since $v_2$ and $v_3$ are neighbors of $v_1$ on $P(n,w)$,
in order for $v_1$ to be a node of a triangle, either the internal edge $(v_2,v_3)$ must be in $T(n,w)$, or
$v_1$ must be part of some internal edge, $(v_1, u)$ in $T(n,w)$. The corollary is proved in the first case, and
also in the second case if $u = v_4$. So assume that $u$ is not $v_4$, and consider the two sub-polygons of $P(n,w)$, call them
$P'$ and $P''$, split
by the edge $(v_1,u)$. Since $P(n,w) = P' \cup P''$ and  $P' \cap P''$ consists only of the edge $(v_1,u)$, $T(n,w)$ 
must consist of edge $(v_1, u)$, and
an optimal triangulation of $P'$,  together with an optimal triangulation of $P''$. Suppose, for concreteness, that $v_4$ is
in $P'$, and note that only one the nodes $\{v_2,v_3\}$ can be in $P'$. Therefore, $v_4$ is the node with the third
smallest weight in $P'$, and by Theorem \ref{tmonotone}, edge $(v_1,v_4)$
must be in any optimal triangulation of $P'$. But that leads to a contradiction, i.e., the assumption that
the optimal triangulation $T(n,w)$ contains neither  $(v_2,v_3)$ nor $(v_1, v_4)$ leads to the conclusion that
$T(n,w)$ must contain $(v_1,v_4)$.
\end{Proof}

\section{A Branching Search-Tree Algorithm}

We describe an $O(n^2)$-time algorithm to solve the polygon triangulation
problem for any monotonic triangle-weight function.
A {\it sub-polygon} of $P(n,w)$ is defined by a subset of nodes of $P(n,w)$,  
{\it ordered} by their relative order on $P(n,w)$. Below, we discuss an algorithm that creates sub-problems, each one
corresponding to a sub-polygon of $P(n,w)$. 

Exploiting Theorem \ref{tmonotone} and Corollary \ref{cYAO}, the algorithm to find an optimal triangulation of
$P(n,w)$ is a {\it top-down} branching-tree algorithm, starting
with the full $P(n,w)$. If at least one of the neighbors of $v_1$ in $P(n,w)$ is not from $\{v_2,v_3\}$, then we apply Theorem
\ref{tmonotone}, creating two or three {\it area-disjoint} sub-polygons, branching on the associated two or three 
sub-problems. The sub-problems do not share any 
area of $P(n,w)$, but do share one or two of the nodes in $\{v_1,v_2,v_3\}$.\footnote{Note that when discussing a sub-problem defined on a sub-polygon $P$,
the symbols $\{v_1,v_2,v_3,v_4\}$ refer to the four least-weight nodes {\it in $P$}
rather than in the original polygon $P(n,w)$.} 
However, if the neighbors of $v_1$ on $P(n,w)$ are $\{v_2,v_3\}$,
then we apply Corollary \ref{cYAO}, branching two ways. One branch further subdivides, creating 
two sub-problems after adding internal 
edge $(v_2,v_3)$; and the other branch subdivides to create two sub-problems after adding internal
edge $(v_1, v_4)$. The algorithm continues from any node in the search tree corresponding to an unsolved sub-problem, 
applying either Theorem \ref{tmonotone}
or Corollary \ref{cYAO}, as appropriate.
The algorithm terminates when each leaf in the search tree corresponds to
a sub-polygon consisting of a triangle.  Each path from the root in the tree defines a set of (non-crossing) internal edges in $T(n,w)$, 
so the path to a leaf defines a full triangulation of $P(n,w)$. 
By Theorem \ref{tmonotone} and Corollary \ref{cYAO}, any minimum-weight triangulation
created by such a path is an optimal triangulation of $P(n,w)$ , and at least one optimal triangulation is found this way.
Later, we will explain how this algorithm relates to the one in \cite{YAO82}.

Branchings created by
application of Theorem \ref{tmonotone} can happen at most $n-1$ times in the full branching tree, 
since each creates {\it area-disjoint} sub-problems. However, the application of Corollary \ref{cYAO} can create two
sub-problems that have intersecting areas, so there could be $\Omega(2^n)$ branchings
created by application of Corollary \ref{cYAO}. This will be reduced to $O(n^2)$ using the key observation from 
\cite{YAO82}.

\subsection{Yao's key observation}

Following the approach in \cite{YAO82}, we next define a {\it bridge} in $P(n,w)$. A bridge is an {\it ordered} 
pair of nodes $(u,v)$ on $P(n,w)$ such that $w(u) < w(x)$, and $w(v) < w(x)$ for every node $x$ strictly 
between $u$ and $v$ on $P(n,w)$,
walking in a {\it clockwise} direction from $u$ to $v$. The nodes on that clockwise walk, together with $u$ and $v$, define a sub-polygon, $P'(u,v)$ of $P(n,w)$. The
remaining nodes, together with $u$ and $v$ define the sub-polygon $P''(u,v)$.
So, a bridge $(u,v)$ splits $P(n,w)$ into the two sub-polygons $P'(u,v)$ and $P''(u,v)$,
each containing both $u$ and $v$. 
By contradiction, it is simple to establish that if $(p,q)$ is another bridge, then either both $p$ and $q$ are in
$P'(u,v)$, or they are both in $P''(u,v)$.\footnote{Note that two bridges may share one or two of their defining nodes, although
both nodes
are shared only by the bridges $(v_1,v_2)$ and $(v_2,v_1)$, i.e., defined by the two least weight nodes on $P(n,w)$.}  That is, bridges of $P(n,w)$ cannot {\it cross}, and therefore there can be at most $n-1$
bridges of $P(n,w)$.

The key insight articulated in \cite{YAO82} is based on the following definition:
A {\it cone} $\CC$ is a sub-polygon of $P(n,w)$ defined either by a bridge $(u,v)$, or a bridge $(u,v)$ and an additional (specific) node $z$
in $P''(u,v)$.
In the first case, cone $\CC$ is the sub-polygon $P'(u,v)$. In the second case, cone $\CC$ is $P'(u,v)$ together with a triangle 
$\{u,v,z\}$
where $z \in P''(u,v)$, and $z$ is the {\it smallest-weight} node in $\CC$.
Note that in the second case, edges $(z,u)$ and $(z,v)$ (which could be internal or external) are part of $\CC$.
Since there can be at most $O(n)$ bridges, and $n$ nodes, the number of
cones is bounded by $O(n^2)$.

We use the notation $(u,v,k)$ to denote a cone $\CC$. The first two values specify the bridge $(u,v)$. The third value, $k$,
is set to 0 if $\CC$ is defined by bridge $(u,v)$ without an added triangle;  and is 
set to $z$ if $\CC$ is defined by $(u,v)$ and triangle $\{u,v,z\}$. Note that a triangle is defined by a {\it set} of
three nodes, while a cone is defined by an {\it ordered list} of three nodes.
The following theorem is from \cite{YAO82} (stated there without proof).

\begin{Theorem}
\label{tconehead}
Every sub-problem generated in the above branching search-tree algorithm is either a triangle (which is a solved
subproblem), or a cone. 
\end{Theorem}

\begin{Proof}
The proof is by induction. For the basis, at the start of the algorithm (and at the root of the search tree) 
there are three cases: Case A) node $v_1$ is 
adjacent to both $v_2$ and $v_3$ in $P(n,w)$; 
Case B) $v_1$ is adjacent to exactly one of $\{v_2, v_3\}$ in $P(n,w)$; Case C) $v_1$ 
is adjacent to neither $v_2$ nor $v_3$ in $P(n,w)$.

In Case A), assume $v_2$ is the neighbor of $v_1$ in $P(n,w)$, in the clockwise direction (the other case is
symmetric). Then, by Corollary \ref{cYAO}, the search tree must branch two ways, adding internal edge $(v_2,v_3)$ to $P(n,w)$ 
on one branch,
and adding internal edge $(v_1,v_4)$ to $P(n,w)$ on the other branch. 
The first branch further subdivides,
creating two sub-problems: one is defined by the triangle $\{v_1,v_2,v_3\}$  (a solved subproblem);
and the other is defined by the bridge $(v_2,v_3)$, and hence is the cone $(v_2,v_3,0)$.
On the other branch, where $(v_1,v_4)$ is added to $P(n,w)$, two sub-problems 
$P'(v_1,v_4)$  and $P''(v_4,v_1)$  are created,
with $v_2 \in P'(v_1,v_4)$  and $v_3 \in P''(v_4,v_1)$. 
Sub-problem $P'(v_1,v_4)$  is defined by
the bridge $(v_2,v_4)$ and the triangle $\{v_1,v_2,v_4\}$, i.e., by the cone
$(v_2,v_4,v_1)$; and sub-problem $P''(v_4,v_1)$ 
is defined by the bridge $(v_4,v_3)$ and the triangle $\{v_1,v_3,v_4\}$, i.e., by the cone
$(v_4,v_3,v_1)$. Note that $v_1$ is the smallest-weight node in $P''(v_1,v_4)$, so the basis holds in Case A.

Case B) In this case, let $x$ be the node in $\{v_2,v_3\}$ that is adjacent to $v_1$ in $P(n,w)$, and let $y$ be the other node
in $\{v_2,v_3\}$.  
Then, by Theorem \ref{tmonotone}, 
internal edge  $(v_1,y)$ will be added to $P(n,w)$. Assume $x$ is in $P'(v_1,y)$ (the opposite case is symmetric). Adding
$(v_1,y)$ creates two sub-problems:
one sub-problem contains node $x$, and is defined by the bridge $(x,y)$ and the triangle
$\{v_1,x,y\}$, i.e., by cone $(x,y,v_1)$, where $v_1$ is the minimum-weight node in the cone. The other sub-problem is defined by the bridge $(y,v_1)$, i.e., by cone
$(y,v_1,0)$. 

Case C) In this case, by Theorem \ref{tmonotone}, internal edges $(v_1,v_2)$ and $(v_1,v_3)$ will be added to $P(n,w)$. 
Assume $v_2$ precedes $v_3$ in a clockwise walk from $v_1$. So adding the edges
creates three
new sub-problems: one is defined by the bridge $(v_1,v_2)$; one is defined by the bridge $(v_3,v_1)$; and one is defined
by the bridge $(v_2,v_3)$ and the triangle $\{v_1,v_2,v_3\}$. These correspond to cones $(v_1,v_2,0)$, $(v_3,v_1,0)$,
and $(v_2,v_3,v_1)$.

Inductively, assume that the theorem holds to some point in the algorithm, and consider expanding an exposed node $h$ in the search
tree, corresponding to a sub-problem defined by  a cone $\CC$. When expanded, 
node $h$ becomes an internal node in the search tree, 
whose children are associated with the sub-problems created from $\CC$. The proof of the inductive step is again by case analysis based
on whether $\CC$ is defined by a bridge only, or by a bridge and a triangle; and by the adjacencies of the smallest-weight
node in $\CC$ with the next two smallest-weight nodes in $\CC$. 
The analyses are similar to those done for the basis,\footnote{Note, however, that Case C) can only occur in the basis.}
but have some subtle differences, and are discussed here for completeness.

Suppose $\CC$ is defined by bridge $(u,v)$ and triangle $\{u,v,z\}$, so node $z$ is the smallest-weight node in $\CC$,
and nodes $\{u,v\}$ are the second and third smallest-weight weight nodes in $\CC$.  Hence, Corollary \ref{cYAO} applies,
and vertex $h$ is expanded in two ways. On one branch, internal edge $(u,v)$ is added to $\CC$, creating two  subproblems:
one defined by the
cone $(u,v,0)$; and one defined by the (solved) subproblem specified by the triangle $\{u,v,z\}$.
On the second branch, internal edge $(z,x')$ is added to $\CC$, where $x'$ is the fourth smallest-weight node in $\CC$. Note
that $(z,x')$ is not a bridge, but $(u,x')$ and $(x',v)$ are bridges. The two subproblems created be adding
edge $(z,x')$ are defined by the cones $(u,x',z)$ and $(x',v,z)$.

If $\CC$ is defined by bridge $(u,v)$ alone, suppose $w(u) < w(v)$ (the opposite case is symmetric).
Let $x$ be the third smallest-weight node on $\CC$. There are two subcases: $x$ is a neighbor of $u$ on $\CC$ 
(and hence on $P(n,w)$), and $x$ is not a neighbor of $u$ on $\CC$. In the first subcase,
Corollary \ref{cYAO} applies, and vertex $h$ branches two ways. On one branch, internal edge $(x,v)$ is added to $\CC$,
creating the solved subproblem of triangle $\{u,v,x\}$, and creating the subproblem defined by the cone $(x,v,0)$.
On the second branch, internal edge $(u,x')$ is added to $\CC$, where $x'$ is the fourth smallest-weight node in $\CC$. 
The two subproblems created be adding edge $(u,x')$ are defined by the cones $(x,x',u)$ and $(x',v,u)$.
In the second subcase, Theorem \ref{tmonotone} applies, and internal edge $(u,x)$ is added to $\CC$, creating two
subproblems: one defined by the cone $(u,x,0)$, and the other defined by the cone $(x,v,u)$.
This completes the induction step.
\end{Proof}

Since there are at most $O(n^2)$ cones in $P(n,w)$, and only $O(n^2)$ triangles where one side is a bridge, we have:

\begin{Corollary}
There are at most $O(n^2)$
distinct sub-problems encountered in the branching search-tree algorithm.
\end{Corollary}

\section{Implementing the search-tree algorithm in $O(n^2)$ time}

The top-level idea for solving the polygon-triangulation problem, for any monotone triangle-weight function, 
is to use the branching search-tree algorithm along with the well-known technique of
{\it memoization} \cite{CLR}. That is, follow a {\it depth-first} expansion of the branching search-tree, and 
whenever a sub-problem is solved, store the value of the solution in a data-structure, $D$, indexed by
a description of the cone associated with the sub-problem. Then, whenever
the search-tree algorithm considers expanding a vertex in the search-tree, to begin the solution of a sub-problem
associated with a cone, 
it checks $D$ to see if that sub-problem has already been solved. If the sub-problem has been solved, the algorithm
backs up from the node, rather than solving the sub-problem again. This approach never expands
two nodes of the search-tree representing the same sub-problem.\footnote{The depth-first 
implementation is required for this consequence,
a fact that is sometimes omitted in textbook discussions of memoization.} 
Hence, the search-tree has at most $O(n^2)$ internal
nodes, and each node (other than the root) has at most two children, so the number of leaves is $O(n^2)$. Each leaf $h$
is either associated with a sub-problem that has already been solved when node $h$ is generated; or is associated with
a sub-problem that is a triangle, whose ``solution" is just the weight of the triangle, 
given by the triangle-weight function.

\paragraph*{Constant-time operations are needed}
Since the number of nodes in the search-tree is bounded by $O(n^2)$, in order to implement the algorithm in that 
time, we must show that each described step of the algorithm can be implemented
in {\it constant} time. These steps are:

1) Inserting into $D$ the solution to a sub-problem described by cone $\CC$, indexed by a description, $(u,v,k)$, of 
$\CC$.

2) Using $D$ to determine if a sub-problem described by cone $\CC$ has already been solved, and if it has been, extracting the
weight of the solution from $D$.

3) If Step 2 determines that the sub-problem described by cone $(u,v,k)$ must be solved, 
apply as appropriate either Theorem \ref{tmonotone} or Corollary \ref{cYAO}, adding a new internal
edge into $\CC$, and expanding a vertex of the search-tree.

\paragraph*{Data structure $D$}

We use a three-dimensional $n \times n \times (n+1)$ array $D$ to represent the encountered cones, where the first two dimensions represent
the ordered pair of nodes defining the bridge $(u,v)$ of the cone, and the third dimension representing, if there is one, 
the additional node, $k$, creating the triangle of the cone. 
The $(u,v,k)$ cell in $D$ holds the weight of the solved sub-problem defined by
the cone $(u,v,k)$. Index $k$ has value 0 if there is
no triangle part of the cone, and otherwise, $k$ specifies the third node of the triangle $\{u,v,k\}$.
Entering and retrieving values from $D$ can be done in constant time, addressing steps 1) and 2).
However, there is a subtle issue
in using $D$ to implement an $O(n^2)$-time algorithm. 

When the algorithm considers expanding a node representing
a sub-problem defined by the cone $(u,v,k)$, the algorithm examines the value of cell $(u,v,k)$ in $D$ to
determine if that sub-problem has already been solved. So, the cells of $D$ must have a special {\it initial} 
value to indicate
that sub-problem $(u,v,k)$ has {\it not} been solved; otherwise, it is possible that whatever initial value is in
$D(u,v,k)$ might lead the algorithm to decide incorrectly that the sub-problem has already been solved. 
The need to initialize $D$ raises a problem:
Although the algorithm can allocate $\Theta(n^3)$-size memory in constant time, it cannot initialize all the
cells in $O(n^2)$ time. The solution to this problem comes from observing that only $O(n^2)$ cells of $D$ will ever be accessed in the
execution of the algorithm, since there are only $O(n^2)$ cones. So, before the search-tree algorithm begins,
we will find the $O(n)$ bridges in $O(n^2)$ time. Then, for each bridge $(u,v)$ of $P(n,w)$, we initialize cells $(u,v,k)$ in $D$, where $k$ ranges from 0 to $n$,
with a value that cannot be confused for a solution to sub-problem.

\paragraph*{Finding all bridges in $O(n^2)$ time}

We know that there are only $O(n)$ bridges, but we will use an $n$-by-$n$ array $S$ to record the bridges and related
information.
From each node $u$ on $P(n,w)$,  walk clockwise around $P(n,w)$ keeping track of the smallest-weight node, other than $u$,
encountered on the walk. Let variable $s(u)$ record that node.  
If a node $v$ with weight smaller than the weight of
$s(u)$ is encountered, $(u,v)$ must be a bridge, at which point the algorithm will initialize $D(u,v,k)$ for $k$ from
0 to $n$; then, record node $s(u)$ and its weight in cell $(u,v)$ of array $S$; 
and set $s(u)$ to $v$. The algorithm ends the walk from $u$ if $w(v) < w(u)$; or when the walk returns
to $u$.  Clearly, every bridge $(x,y)$ will be detected on a walk starting from node $x$, so the $n$ walks will find all of the bridges,
and set the values in $D$ and $S$, in $O(n^2)$ time. 

The array $D$ is used here to establish the $O(n^2)$ {\it worst-case} running time, 
but for practical implementation, we avoid the use of 
$D$ by {\it hashing}
the descriptions of the cones created during the search-tree algorithm. The use of hashing invalidates the $O(n^2)$ bound, but
is very effective in practice. Further, without $D$
we only need to initially find the bridges, and not initialize $D$,
in which case, the walks can be combined into a faster algorithm that finds all of the bridges, with an $O(n)$
worst-case running time. See Procedure MarkBriges in \cite{YAO82} for details. That algorithm can also be
modified to set $S(u,v)$ for each of the $O(n)$ bridges, again in $O(n)$ total time.

\paragraph*{Implementing each Step 3 in constant time}

Let $(u,v,k)$ describe cone $\CC$ representing a subproblem to be solved.  
We must implement Step 3 in constant time. Given the operations detailed in the proof of Theorem \ref{tconehead}, the only
remaining issue is to identify the third and (sometimes) the fourth smallest-weight node(s) on $\CC$. We show 
that this only takes constant time using array $S$.
We divide the argument into the case that $k = 0$, and the case that $k > 0$.

Suppose first that $k = 0$, so $\CC$ is defined by the bridge $(u,v)$. Suppose $w(u) < w(v)$ (the opposite case
is symmetric), and let $x$ be
the neighbor of $u$ in the clockwise direction on $P(n,w)$, and hence also on $\CC$. 
Then extract $S(u,v)$, which will be the third smallest-weight node on $\CC$. If 
$x \neq S(u,v)$, then Theorem \ref{tmonotone} applies, and edge $(u,S(u,v))$ is added to $\CC$. 
If $x = S(u,v)$, then Corollary \ref{cYAO} applies, and we need to identify the fourth smallest-weight
node on $\CC$. But when $x$ is the third smallest-weight node on $\CC$ (and $u$ and $v$ are the first and second smallest-weight
nodes on $\CC$), edge $(x,v)$ is a bridge, so the fourth smallest-weight node on $\CC$ is  given by $S(x,v)$.

Now suppose that $k > 0$, so $\CC$ is described by the bridge $(u,v)$  and the triangle $\{u,v,k\}$. Then, $k$ is the
smallest-weight node on $\CC$, and $\{u,v\}$ are the second and third smallest-weight nodes on $\CC$. In that case,
Corollary \ref{cYAO} applies, and node $S(u,v)$ will be the fourth smallest-weight node on $\CC$.

\paragraph*{Summary of the time used}

Using the three-dimensional array $D$, we have shown all of the details needed to establish the $O(n^2)$ time bound, including the
time to initialize the needed parts of $D$. When hashing is used to avoid the need for $D$, the running time is dominated by the time
for hashing, and by the number of cones created
in the execution of the search-tree algorithm, which (as detailed in Section \ref{sempirical}) can be considerably less than the total number of cones in $P(n,w)$.

\paragraph*{Yao's Algorithm}

The algorithm in \cite{YAO82} essentially finds all of the cones in $P(n,w)$ and builds a partial order, ordering the cones  by area-inclusion.  
Cone $\CC'$ precedes cone $\CC$ in the partial order if the sub-polygon defined by $\CC'$ is completely contained in the sub-polygon defined
by $\CC$. A dynamic program solves the polygon triangulation problem for each cone $\CC$, bottom up in the partial order. That is,
the triangulation problem for a cone $\CC$ is solved after all the subproblems defined by cones that precede $\CC$ 
have been solved. Optimal solutions to solved
subproblems are held and extracted from a table, but no details of the table are discussed in \cite{YAO82}. Thus, 
the subtle issue of how
such a table is initialized in $O(n^2)$ time is not addressed in \cite{YAO82}.
The running
time for this algorithm is dominated by the {\it total} number of cones in $P(n,w)$, which is bounded by $O(n^2)$.

\section{The Time-Bound is Tight}

Since the time for Yao's algorithm is dominated by the {\it total} number of cones in $P(n,w)$, we show that the algorithm takes $\Theta(n^2)$ time 
with a construction containing $\Theta(n^2)$ cones. This also establishes that the search-tree algorithm using array $D$ takes $\Theta(n^2)$ time,
since the initialization of $D$ puts values into a set of cells of $D$ that is at least as large as the number of cones of $P(n,w)$.
However, the practical implementation of the search-tree algorithm uses hashing in place of $D$, 
and the running time of that version is dominated by the
number of cones {\it encountered} in the search (assuming that each hashing operation takes constant time). Our construction also establishes
that the search-tree algorithm encounters $\Theta(n^2)$ cones.

Let $P^*(2n, w)$ be a polygon such that the vertices have the form of 
$v_1 - v_2 - v_4 - \dots - v_{2n-2} -  v_{2n} - v_{2n-1} - \dots - v_5 - v_3$ where $v_1 < v_2 < \dots < v_{2n}$. 
We will show that $P^*(2n,w)$ has $\Theta(n^2)$ cones. We first observe that every bridge either has the form 
$b_{2k-1} = (v_{2k}, v_{2k-1})$  or $b_{2k} = (v_{2k},  v_{2k+1})$ for $1\le k \le n-1$. For each bridge $b_i$, 
there is one cone defined by $b_i$ alone (called a \textit{degenerate cone}) and $i-1$ cones defined by bridge $b_i$ and a triangle formed by $(b_i, v_k)$, where $1\le k\le i-1$. Hence, the total number of cones is $\Theta(n^2)$. It follows that Yao's algorithm achieves the tight bound $\Theta(n^2)$. The following theorem will show that the bound is also tight for the branching search-tree algorithm.

\begin{Theorem}
\label{tightbound}
For each bridge $b_i$ where $1\le i\le 2n-2$, every cone defined by $b_i$ is visited by the branching search-tree algorithm. Hence, the algorithm runs in $\Theta(n^2)$ on $P^*(2n,w)$.
\end{Theorem}


\begin{Proof}
The proof is by induction. 
The bridge $b_1 = (v_2, v_1)$ defines only one cone $P^*(2n, w)$ which is at root of the search tree. 
Since $v_1$ is adjacent to both $v_2$ and $v_3$, the search tree must branch two way, adding internal edge $(v_2, v_3)$  on one branch, and adding $(v_1, v_4)$ on the other branch. These are reduced to the two cones defined by $b_2 = (v_2, v_3)$. So, all cones defined by $b_2$ are visited.

Assume the algorithm visits every cone defined by $b_i$ where $i \ge 2$. We will show that it must visit every cone defined by $b_{i+1}$. There are two types of cones defined by $b_{i+1}$.
\begin{itemize}
\item[] Case 1: The cone is the degenerated cone. This cone is visited by the search tree because it is one of the two branches for the degenerated cone of $b_i$ which got visited by our assumption. 
\item[] Case 2: The cone is formed by $(b_{i+1}, v_k)$ where $1\le k \le i$. For $1\le k \le i-1$, the cone $(b_{i+1}, v_k)$ is one of the two branches of the cone $(b_i, v_k)$ which is visited by our assumption. Also, the cone $(b_{i+1}, v_i)$ is one branch for the degenerated cone of $b_i$ and then it is visited by our search tree.
\end{itemize}
Therefore, the branching search-tree algorithm visits all cones. Since there are $\Theta(n^2)$ cones,  the branching 
search-tree algorithm has lower bound running time $\Theta(n^2)$. Therefore, the running time is $\Theta(n^2)$.
\end{Proof}

\section{Approximation Algorithm for the Additive Function}
In \cite{HU-SHING1981}, Hu and Shing gave an $O(n)$-time heuristic algorithm to find a near-optimal partition of any 
convex polygon, using the multiplicative triangle-weight function. The partition that their method finds is guaranteed to 
have weight less than $15.5\%$ more than the optimal triangulation, which is an unusually small error bound for a linear-time
algorithm.  We have not been able to obtain the same result using the {\it additive} triangle-weight function. However,
we give here a modification of the Hu-Shing algorithm that is guaranteed to have weight less than 
$\frac{1}{3}$ more than the optimum, when the additive triangle-weight function is used. The proof of this result is
very similar to the proof of the error bound of $15.5\%$ in \cite{HU82}, so we will only sketch the places
where the two analyses differ. A reader can verify our result by following the proof in \cite{HU82}, making
the changes indicated here.
We begin by stating two theorems from \cite{HU-SHING1981}.

\begin{Theorem} Recall that $v_1, v_2, v_3$ are the three smallest-weight nodes of a convex polygon. 
Assume that $v_1$ is adjacent to both $v_2$ and $v_3$. 
A necessary, but not sufficient, condition for the edge $(v_2, v_3)$ to be connected in optimum partition is:
\[
v_1 + v_4 > v_2 + v_3.
\]
Moreover, if edge $(v_2,  v_3)$ is not in an optimum partition, then edge $(v_1, v_4)$ must be in that optimum partition.
\end{Theorem}

\begin{Theorem}
Let $v_m$ be a node with weight that is greater than both if its neighbors, $v_p$ and  $v_q$. 
Then edge $(v_p, v_q)$ must be an edge in the optimum partition of the polygon.
\end{Theorem}

We next describe our heuristic algorithm.

\begin{algorithm}
\caption{Heuristic\_Optimum\_Polygon$(P(n,w))$}
\begin{algorithmic}[1]
\STATE Let $V_1, V_2, \dots, V_n$ be the vertices of $P$ in clockwise ordering where $w(V_1) = v_1$ is the smallest vertex.
\STATE $stack \rightarrow [V_1, V_2, V_3]$
\FOR {$v$ from $V_4$ to $V_n$}
\STATE $x = stack.pop()$ 
\STATE $y = stack.top()$
\IF{ $\vert stack \vert\ge 2$ and $y + v < v_1 + x$}
\STATE Join the edge ($v, y$)
\ELSE
\STATE $stack.push(x)$
\STATE $stack.push(v)$ 
\ENDIF
\ENDFOR

\WHILE{ $\vert stack \vert \ge 3$}
\STATE $x = stack.pop()$ 
\STATE $y = stack.top()$
\STATE Join the edge $(v_1, y)$
\ENDWHILE
\end{algorithmic}
\end{algorithm}

Let $P$ be the smallest polygon that gives the worst error ratio for the heuristic algorithm. We know that $P$ must consist of 
five vertices or more because the heuristic algorithm always finds the optimum partition for any quadrilateral. 

\begin{Theorem}
Polygon $P$ must be a strictly monotone polygon, i.e., the vertices of $P$ must have one of the following forms:
\begin{itemize}
\item[i.] $v_1 - v_2 - v_4 - \dots - v_{n-2} - v_n - v_{n-1} \dots - v_5 - v_3$ if $n$ is even, or
\item[ii.] $v_1 - v_2 - v_4 - \dots - v_{n-1} - v_n - v_{n-2} - \dots - v_5 - v_3$ if $n$ is odd.
\end{itemize}
Moreover, the maximum error is achieved when all the vertices have two kinds of weights,
\[
v_{n-2} = v_{n-3} = \dots = v_2 = v_1 = x\ge 1
\]
and 
\[
v_{n-1} = v_n = tx,
\]
where $t\ge 1$.
\end{Theorem}

The proof of this theorem is very similar to the proof of Theorem $8$ in \cite{HU-SHING1981}.  
Given the weights of $P$ as describing in Theorem $5.3$, we can easily find the cost of optimum partition and the cost of 
heuristic partition. Connecting $v_i-v_{i+1}$ for all $2\le i \le n-1$ is an optimum triangulation. However, the 
heuristic algorithm returns the triangulation where $v_1$ is connected to all other vertices. 
Let $C$ be the cost of the partition given by the heuristic algorithm, and $C_{opt}$ be the cost of the optimal partition. Then, 
\begin{align*}
C &= (2t+1)x + 2(t+2) x + 3(n-5)x\\
C_{opt} & = (2t+1)x + (t+2) x + 3(n-4) x
\end{align*}
The error bound is given by,
\[
E = \dfrac{C - C_{opt}}{C_{opt}} = \dfrac{t-1}{3t + 3 + 3(n-4)}
\]
The error bound is maximum when $n = 5$. Therefore, 
\[
E = \dfrac{t-1}{3(t+2)} < \dfrac{t+2}{3(t+2)} = \dfrac{1}{3}
\]
The upper bound is tight because $\lim\limits_{t\rightarrow \infty} E = \frac{1}{3}$, i.e., 
for any $\varepsilon > 0$, we can find a value of $t$ such that $E = \frac{1}{3} - \varepsilon$.

\section{Empirical Results With Random Data}
\label{sempirical}

We now compare our branching search-tree algorithm with Yao's algorithm on random data. All the experiments are conducted on an Intel 2GHz Core i7 CPU. The codes can be found in the following link,
\[
\text{\href{https://github.com/thongle91/Speedup-Matrix-Chain-Multiplication}{https://github.com/thongle91/Speedup-Matrix-Chain-Multiplication}}
\]

For each value of $n$, we generated several hundred different sets of weights for vertices of polygon $P(n)$,
and then ran both algorithms, and the classic $O(n^3)$-time dynamic programming method,  on each dataset. 
Table \ref{tab:table1} shows the average running times of these algorithms for each dataset,
establishing that the
branching search-tree algorithm performs much better on random data than does Yao's algorithm, and dramatically better than the classic, widely-taught,  dynamic programming
method.

\begin{table}[H]
  \centering
  \caption{Experiment results}
  \label{tab:table1}
  \begin{tabular}{|c|c|c|c|c|c|c|}
    \toprule
      & $n = 100$ & $n = 500$ & $n = 1000$ & $n = 5000$ & $n = 10000$ & $n = 10^5$ \\
    \midrule
    BST Algorithm & $0.0001s$ & $0.01s$ & $0.02s$ & $0.1s$ & $0.4s$ & $6s$\\
    \midrule
    Yao's Algorithm & $0.01s$ & $0.4s$ & $1.3s$ & $44s$ & $631s$ & $> 1hr$ \\
    \midrule
    Classic DP $O(n^3)$ & $0.14s$ & $24.5s$& $246s$ & $> 1hr$ & $> 1hr$& $> 1hr$ \\
    \bottomrule
  \end{tabular}
\end{table}

We also counted the number of visited cones for each dataset  in both the BST algorithm and Yao's algorithm. 
The results are shown in the following two graphs. We observe that the number of visited cones in the 
branching-search tree algorithm is empirically  $\Theta(n\log n)$, while it is $\Theta(n^2)$ for Yao's algorithm.

\begin{figure}[h]
\includegraphics[scale=0.4]{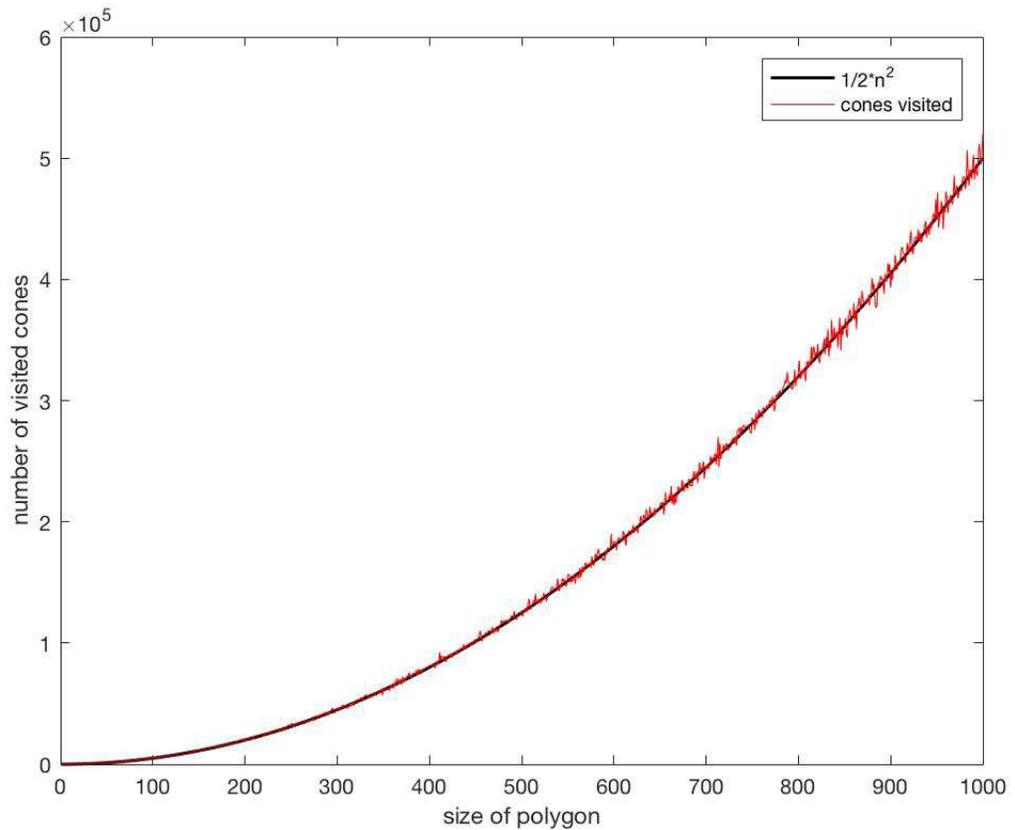}
\caption{The number of visited cones in Yao's algorithm.}
\end{figure}

\begin{figure}[h]
\includegraphics[scale=0.40]{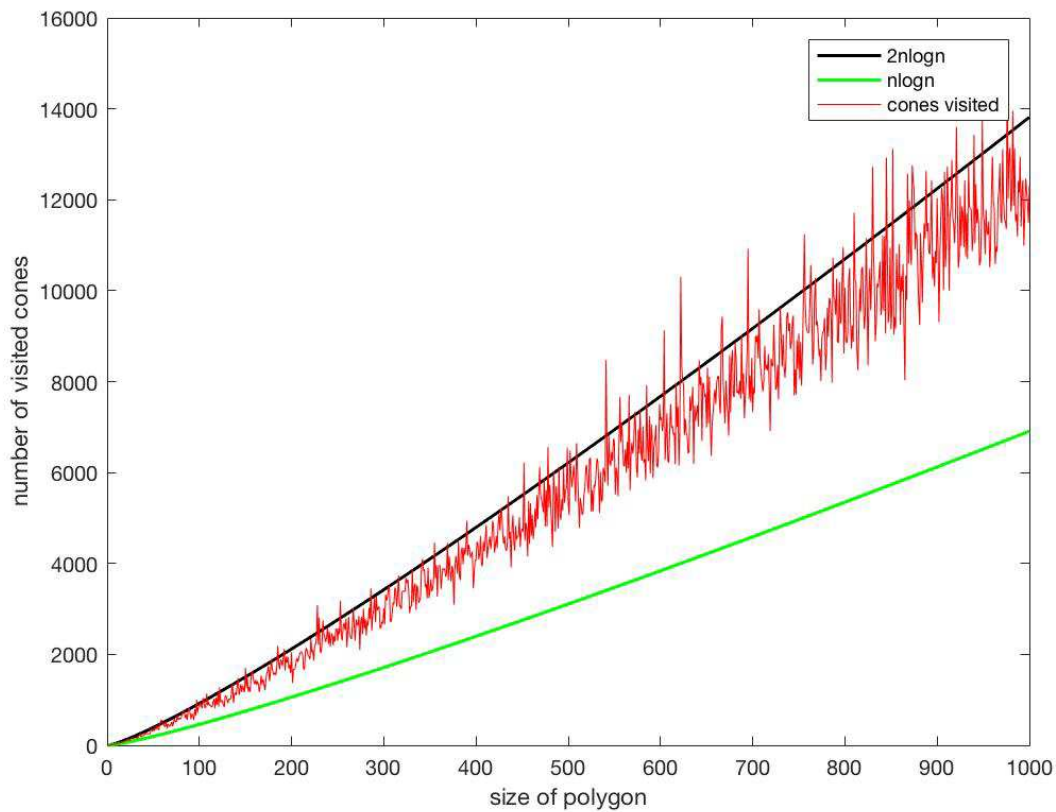}
\caption{The number of visited cones in branching-search tree algorithm.}
\end{figure}

\end{document}